\documentclass{aa}
\usepackage{txfonts}
\usepackage{graphicx}
\begin{document}
 
\title{Degenerate sterile neutrino dark matter in the cores 
of galaxies}

\author{Faustin Munyaneza\thanks{Humboldt Fellow}\inst{1}  \and Peter L. Biermann\inst{1,2,3}}
\offprints{ F. Munyaneza, \email{munyanez@mpifr-bonn.mpg.de}}
\institute{Max-Planck-Institut  f\"ur Radioastronomie,
         Auf dem H\"ugel 69,
         D-53121 Bonn, 
         Germany
         \and Department of Physics and Astrononomy, University of Bonn, Germany
         \and Department of Physics and Astronomy,
              University of Alabama, Tuscaloosa, AL, USA}
         
\date{Received <date>/ Accepted <date>}

\abstract{
We study the distribution of  fermionic dark matter at the center of galaxies
using  NFW, Moore and isothermal density profiles 
and show that dark matter becomes degenerate 
for  particle masses of a few {\rm keV}
and for  distances less than
 a few 
parsec from the center of our galaxy. 
A compact degenerate core forms after galaxy merging and boosts the 
growth of supermassive black holes
at the center of galaxies.
To explain the galactic center black hole of mass of $\sim 3.5 \times 10^{6}M_{\odot}$ and
a supermassive black hole of $\sim 3 \times 10^{9}M_{\odot}$ at a redshift
of 6.41 in SDSS quasars, we require
a degenerate core of mass between $3 \times 10^{3} M_{\odot}$
and $3.5 \times 10^{6}M_{\odot}$. This constrains  
  the mass of the dark matter particle
 between   $0.6 \ {\rm keV}$ 
 and $82~{\rm keV}$.
 The lower limit on the dark matter mass 
  is improved to ~{\rm 7~keV} if exact solutions of Poisson's equation
  are used in the isothermal power law case.
 We argue that the constrained particle could be the long sought 
dark matter of the Universe that is 
interpreted here as  
  a sterile neutrino.
}
\titlerunning{Constraints on the dark matter particle mass}
\authorrunning{F. Munyaneza \& P.L. Biermann}
\maketitle

\section{Introduction}
Precision observations of the cosmic microwave background and of large scale structure
confirm the picture in which 96\% of the matter density of the Universe is made of dark
energy and dark matter (DM), which could
be revealed only by their gravitational interaction (Spergel  et al. \cite{spergel03}).
The nature of these two forms of matter is still unknown.
Many candidates have been proposed for DM. These include Cold Dark Matter (CDM) particles of masses
heavier than $1 {\rm GeV/c^{2}}$ (Bertone et al. \cite{bertone05}) and Warm Dark Matter (WDM)
 particles such as 
sterile neutrinos (Dodelson \& Widrow \cite{dw94}).
Recently, there has been a renewed interest in sterile neutrinos as candidates for DM as they could be a
 natural extension of 
the minimal  standard model (MSM) of electroweak interactions (Weinberg \cite{weinberg67}; Glashow \cite{glashow61}).
One of these sterile neutrinos could be the DM, while the other two could help explain baryogenesis
(Asaka \& Shaposhnikov \cite{asaka05}; Shaposhnikov \cite{shapo06}). 

Numerical cosmological N-body simulations suggest DM density profiles
which follow  $\rho \sim r^{-\gamma}$ law, with 
$\gamma \approx 3$ in the outer parts of the halos
and $1 \stackrel {\textstyle <}{\sim}   \gamma  \stackrel {\textstyle <}{\sim} 2$ 
inside  a few kpc (Navarro et al. \cite{nfw97}; Moore et al. \cite{moore98}, \cite{moore99};
Klypin  et al. \cite{klypin02} and Power et al. \cite{power03}).
Although the Milky Way is
well studied in the inner 3 - 10 kpc region,
little is known about the DM distribution on smaller scales, i.e. 
$r \stackrel {\textstyle <}{\sim}0.1 \ {\rm pc}$ where 
the 
gravitational potential of baryons dominates over DM.
In addition, there is a mounting evidence that  a black hole of mass
$M \sim 3.5 \ 10^{6}M_{\odot}$ dominates  the mass distribution in the inner one parsec of the Galaxy
(Sch\"odel et al. \cite{schodel03}; Ghez et al. \cite{ghez05}).
Thus the investigation of the  distribution of DM around 
the black hole in the Galactic center 
is important as it could give some hints on the nature of the  DM particle
(Gnedin \& Primack \cite{gnedin04};  Bertone \& Merritt
\cite{bm05}).
   Recently, we  have established that the  inner DM density profile scales as  
 $\rho \sim r^{-3/2}$ under the assumption  that the black hole feeds from degenerate 
 fermionic
  DM (Munyaneza \& Biermann \cite{mb05}).
  It was then shown
    that a DM particle of order of $m_{s} \sim 10$ keV 
    could explain the growth 
    of supermassive black holes of $10^{6}$ to $10^{9}M_{\odot}$
from stellar seed black holes.
It is interesting to note that a  sterile neutrino mass
in overlapping mass range,  with a small mixing angle with the active neutrinos,
  has  also been suggested 
by Kusenko (\cite{kusenko04}) to explain the high velocities of pulsars at birth in supernovae.
Moreover, a DM particle mass in the 
 keV range  can be constrained 
 from  X-ray background 
studies (Drees \& Wright \cite{dw00};  Abazajian et al.  \cite{abaza01} and Dolgov \& Hansen \cite{dh02}).
Here, it is worth mentioning that  an upper limit of  a sterile neutrino mass
of $6.3 \ {\rm keV}$ from X-ray background seems to be in conflict with the lower limit
of $14 \ {\rm keV}$ from SDSS Lyman-alpha forest (Abazajian \& Koushiappas \cite{ak06}).
However, 
the non-thermal phase distribution of the DM particles has the potential to modify
the Lyman alpha limits, while leaving the X-ray limits intact, so eliminating the contradiction
with this additional degree of freedom.
Depending on the specific model for the production of the DM particles, their initial phase
space distribution is possibly largely sub-thermal.
Recently, Biermann \& Kusenko (\cite{bk06}) established that 
the decay of a such sterile neutrino could speed up
the formation of molecular hydrogen and boost the early
star formation and reionisation in agreement with the  WMAP 3-year results
(Spergel et al.  \cite{spergel06}).

The purpose  of this Letter is to study the constraints on the DM particle mass
in the central region of the galaxy. Given that there is still uncertainty about the DM
profile around the central black hole, we will assume  
standard Navarro-Frenk-White (NFW), Moore or 
isothermal gas sphere 
profiles and investigate what happens
 if the  DM particles
 become degenerate and form a degenerate 
dark matter star also  called  a fermion 
ball (Munyaneza \& Viollier \cite{mv02}; Munyaneza \& Biermann \cite{bm05}).
 We  then get the lower limits on the DM particle mass by assuming that
the mass of the fermion ball  cannot be more than the 
mass of  $3.5 \times 10^{6}M_{\odot}$ for the Galactic center
black hole. An upper limit to the DM mass will be established from the condition that the mass of the fermion ball
should be greater than  about $3 \times 10^{3}M_{\odot}$. The choice of this mass
comes from  
the growth mechanism arguments
 to form supermassive black holes of mass of $10^{9}M_{\odot}$ 
in SDSS quasars (Munyaneza \& Biermann \cite{mb05}).
Moreover, 
  a mass of $3 \times 10^{3} M_{\odot}$ is the
  upper limit of masses for which there is no black hole
  in galaxies such as M33 (Gebhardt et al. \cite{geb01}, \& Barth et al. \cite{barth05}). 
Thus the Galactic center provides us with a  fertile testing ground for 
DM theories and cosmological structure evolution.

\section{Degenerate cores}
We study the  following density profiles obtained through galaxy mergers
\begin{equation}
\rho(r) = \rho_{0} \frac{1}{\left(\frac{r}{r_{0}}\right)^{s}}  
\frac{1}{\left(1 + \frac{r}{r_{0}}\right)^{3-s}},
\label{eq:1}
\end{equation}
with $s < 3$ and $\rho_{o}$ is the characteristic density at a size of $r_{0}= 3 \ {\rm kpc}$. 
 The value of $\rho_{0}$ will be found from the 
 constraint that the DM mass enclosed  at $r= 3 \ {\rm kpc}$ is 
about $10^{10} M_{\odot}$  for the Milky Way galaxy (Klypin et al. \cite{klypin02}).
The index $s$ characterizes the density power law slope and we  will
use the value of $s=1$ for NFW profile (Navarro, Frenk \& White \cite{nfw97}), $s=1.5$ for Moore et
 al. (\cite{moore98}, \cite{moore99})
 profile and 
$s=2$ for the isothermal gas sphere case. 
The full distribution function corresponding to the
the density of equation (\ref{eq:1}) can be found in Hansen et al. (\cite{hansen05}). 
Of course, this distribution function  does not allow for  the Fermi 
Dirac nature of the DM particles.
The mass enclosed within a radius $r$ and for 
small $r \stackrel {\textstyle <}{\sim} r_{0}=3$ kpc
is given by
\begin{equation}
M(r) = \int_{0}^{r} 4 \pi r^{2} \rho(r) dr=
  \frac{4 \pi \rho_{0}}{3-s} r_{o}^{3} \left(\frac{r}{r_{0}}\right)^{3-s}.
\label{eq:3}
\end{equation}
 The rotational velocity $v_{rot}$ corresponding to the mass enclosed $M(r)$ 
is 
 \begin{equation}
 {\rm v}_{rot} =\sqrt{\frac{GM(r)}{r}}= \left(\frac{4\pi G \rho_{0}}{3-s}
  \right)^{1/2} r_{0} \left(\frac{r}{r_{0}}\right)^{1-s/2} .
 \label{eq:5}
 \end{equation}
   Here, we will consider only DM contribution to ${\rm v}_{rot}$.  
 The numerical simulation results of inner density power law density distributions suggest that
 for a large number of particles, their kinetic energy is strongly reduced and the more so, 
 the deeper
 they are near the center. The central particle populations are essentially cooled 
 in the core in the violent relaxation of the merger, with  a local temperature $T \sim r^{2-s}$.
   The  
  degeneracy condition can be written as
\begin{equation}
\left(\frac{g_{f}}{6\pi^{2}}\right)^{1/3}\frac{m_{s}{\rm v}}{n^{1/3}} = \hbar \, .
\label{eq:6}
\end{equation}
Here,  $m_{s}$ is the DM particle mass to be constrained,
$n = \rho/m_{s}$ is the DM particle number density,
$\hbar$  is the Planck's constant and $g_{f}$ is the spin degree of freedom, i.e.
$ g_{f}=2$ for Majorana and $g_{f}=4$ for Dirac's particles.
The constraint (\ref{eq:6}) implies a degenerate Fermi
Dirac distribution which of course does not have
 a long power law high energy tail.
The Fermi Dirac distribution function $f_{FD}$ is given by 
 \begin{equation}
 f_{FD}(E) = K \frac{1}{{\rm exp} (\frac{E-\mu}{kT}) + 1}, \, \, \, \, \, \, \, K = N \left(\int_{-\infty}^{0} dE \frac{1}{{\rm exp} (\frac{E-\mu}{kT}) + 1} \right)^{-1},
 \label{fd1}
 \end{equation}
 where $E$ is the DM total energy, $\mu$ the chemical potential, $T$ the temperature and
 $ K $ is the normalisation constant and $ N$ the total number of DM particles.
  For high energies, i.e. $E-\mu >> kT$, the Fermi - Dirac distribution function
 becomes Maxwellian, i.e. $f_{DE} \sim {\rm e}^{-(E-\mu)/kT}$
   whereas direct numerical simulations
 favour a velocity distribution with a long power law high 
 energy tail (Hansen \& Stadel \cite{hs06},
 Hansen et al. \cite{hansen05}) with the following profile
 \begin{equation}
 f({\rm v}) =\left[1-(1-q) \left(\frac{ {\rm v}}{\kappa_{1} \sigma}
 \right)^{2}\right]^{\frac{q}{1-q}},
 \label{eq:hs}
 \end{equation}
 where $q$ and  $\kappa_{1}$ are free parameters. The latter distribution becomes
  Maxwellian for $q=1$. We also note that for ${\rm v} \rightarrow 0$, the distribution function
  given by equation (\ref{eq:hs})
  becomes indistinguishable from a degenerate Fermi Dirac distribution 
  at absolute zero Temperature.
  To reproduce the characteristic features 
  of the distribution given by equation (\ref{eq:hs}) 
  in the Fermi Dirac distribution, we 
 introduce 
 a multi-temperature distribution function as follows
 \begin{equation}
 f(E) = K \sum_{1}^{\infty} \frac{1}{{\rm exp} (\frac{E-\mu_{n}}{kT_{n}}) + 1},
 \label{eq:fd1}
 \end{equation}
 where $T_{n} = 2^{n} T_{0} $ and the chemical potential $\mu_{n}$ is choosen from the 
 condition that ${\rm exp}\left(\frac{\mu}{k 2^{n} T_{0}}\right) = \frac{1}{p^{n}}$. In result we
 get
  \begin{equation}
  \mu_{n} = -kT_{0} n 2^{n} \ln p,
  \end{equation}
  where $n$ is the power index and $p^{n}$ shows by how much the distribution function has been
  lowered. 
 In the limit of $n\rightarrow \infty$, the sum from equation (\ref{eq:fd1}) can be converted into
 the following integral 
   \begin{equation}
  f(E) = K \int_{T0}^{\infty} \frac{1}{{\rm exp} (\frac{E-\mu}{kT}) + 1} \frac{dT}{T},
 \label{eq:fdm}
  \end{equation}
  Thus, the chemical potential and  the temperature parameters allow us to achieve 
  a high  energy
  power law tail in the Fermi Dirac distribution.
  For realistic velocity distributions  obtained from large scale simulations such as given
  by equation (\ref{eq:hs}), the dominant contribution to the number of particles
  is still the basic temperature given by $\kappa_{1} \sigma$,
   demonstrating that a Fermi Dirac
  distribution is a viable first approximation. The modification to the
   degeneracy criterion
  of equation (\ref{eq:6}) is small. In addition, the density powerlaw slope $s$ in NFW, Moore
  and isothermal distributions can be related to our parameter $p$ so that the number of particles at the
  temperature $T_{0}$ dominates by far over all other partial sums.

  The characteristic Fermi velocity is denoted by ${\rm v}$ and  will be taken here as the
 rotational velocity, ${\rm v}_{rot}$.
 Using equations (\ref{eq:1}) and  (\ref{eq:5}), we find that
 the degeneracy condition (\ref{eq:6}) becomes 
 fulfilled for particle masses of a few {\rm keV} and for distances 
   less than  a few parsec.
 Here, we point out that the physics of degenerate fermion balls that obey equation (\ref{eq:6})
 was studied in a series of papers (Bili\'c et al. \cite{bmv99}; Tsiklauri \&  Viollier \cite{tv98}; 
 Munyaneza et al.  \cite{mtv98}, \cite{mtv99}; Munyaneza \& Viollier \cite{mv02}).

Introducing the velocity dispersion $\sigma^{2} =\frac{GM_{0}}{2r_{0}}$, where
 $M_{0}=\frac{4\pi r_{0}^{3}\rho_{0}}{3-s}$ is the total
mass of DM contained within a size of $r_{0}=3 \ {\rm kpc}$.
  As both  the velocity ${\rm v}$ and the number density $n$ depend on the radial coordinate, 
 the degeneracy condition (\ref{eq:6})  allows to get  a 
constraint on the DM particle mass $m_{s}$ and  the radius $r$.
Let us denote by $R_{DM}$ the radius at which DM becomes degenerate
\begin{equation}
  R_{DM} = \left(\frac{6\pi^{2}}{g_{f}}\right)^{\frac{2}{6-s}}
 \left( \frac{3-s}{4\pi}\right)^{\frac{2}{6-s}}
 2^{\frac{-1}{6-s}}
 l_{{\rm pl}}
 \left(\frac{c}{\sigma}\right)^{\frac{2}{6-s}}
 \left(\frac{l_{{\rm pl}}}{r_{0}}\right)^{\frac{s-2}{6-s}}
 \left(\frac{ m_{{\rm pl}}}{m_{s}}\right)^{\frac{8}{6-s}} .
 \label{eq:7}
 \end{equation}
 Using the last expression for the radius, we can calculate the mass $M_{DM}$
  of DM enclosed 
 within  the above radius $R_{DM}$ to be
 \begin{equation}
 M_{DM}= 
  \left(\frac{6\pi^{2}}{g_{f}}\right)^{\frac{6-2s}{6-s}}
 \left( \frac{3-s}{4\pi}\right)^{\frac{6-2s}{6-s}}
 2^{\frac{3}{6-s}}
 m_{{\rm pl}}
 \left(\frac{\sigma}{c}\right)^{\frac{6}{6-s}}
 \left(\frac{l_{{\rm pl}}}{r_{0}}\right)^{\frac{3(2-s)}{6-s}}
 \left(\frac{m_{{\rm pl}}}{m_{s}}\right)^{\frac{8(3-s)}{6-s}}, 
 \label{eq:8}
 \end{equation}
 where $m_{{\rm pl}}= \left(\hbar c/G\right)^{1/2}$ and
   $l_{{\rm pl}}=\left(\hbar G/c^{3}\right)^{1/2}$ are the Planck's mass and length,
    respectively.
    We wish to emphasize that this is just the mass derived
    from the original power law density distribution, and so does not include any additional
    DM accretion to the DM degenerate configuration. 
      We then rewrite   
the equations (\ref{eq:7}), (\ref{eq:8})  for each 
value of the steepness index
$s$.
Thus, for  NFW profile, i.e. $s= 1$ we get
 \begin{equation}
M_{DM} =
 5.24 \ m_{{\rm pl}}
 \left(\frac{\sigma}{c}\right)^{6/5}
 \left(\frac{l_{{\rm pl}}}{r_{0}}\right)^{3/5}
 \left(\frac{m_{{\rm pl}}}{m_{s}}\right)^{16/5},
 \label{eq:09}
 \end{equation}
 \begin{equation}
  R_{DM} = 
  1.62 \ l_{{\rm pl}}
 \left(\frac{c}{\sigma}\right)^{2/5}
 \left(\frac{l_{{\rm pl}}}{r_{0}}\right)^{-1/5}
 \left(\frac{m_{{\rm pl}}}{m_{s}}\right)^{8/5}, 
 \label{eq:10}
 \end{equation}
for the total  mass  and the size of the degenerate core, respectively.
For Moore profile, i.e. $s=1.5$, the total mass $M_{DM}$ and the size $R_{DM}$ 
of the degenerate core are given by
\begin{equation}
M_{DM} =
 3.68 \ m_{{\rm pl}}
 \left(\frac{\sigma}{c}\right)^{4/3}
 \left(\frac{l_{{\rm pl}}}{r_{0}}\right)^{1/3}
 \left(\frac{m_{{\rm pl}}}{m_{s}}\right)^{8/3}, 
 \label{eq:11}    
\end{equation}
 \begin{equation}
  R_{DM} = 
  1.50 \  l_{{\rm pl}}
 \left(\frac{c}{\sigma}\right)^{4/9}
 \left(\frac{l_{{\rm pl}}}{r_{0}}\right)^{-1/3}
 \left(\frac{m_{{\rm pl}}}{m_{s}}\right)^{16/9}, 
 \label{eq:12}
\end{equation}
and finally for the isothermal gas sphere, i.e for $ s=2$, the mass
$M_{DM}$  and the size $R_{DM}$ scale as 
\begin{equation}
M_{DM} =
 2.58 m_{{\rm pl}}          
 \left(\frac{\sigma}{c}\right)^{3/2} 
 \left(\frac{ m_{{\rm pl}}}{m_{s}}\right)^{2}, \, \, \, R_{DM} = 
  1.29l_{{\rm pl}}
 \left(\frac{c}{\sigma}\right)^{1/2}
 \left(\frac{ m_{{\rm pl}}}{m_{s}}\right)^{2} . 
      \label{eq:13}
    \end{equation}
In order to investigate the behaviour of the degenerate DM mass $M_{DM}$ and 
its size $R_{DM}$ as functions of the DM mass $m_{s}$, we 
fix the total DM mass $M_{0} = 10^{10}M_{\odot}$ at a distance of $r_{0}= 3 \ {\rm kpc}$ and this gives
a velocity dispersion  of $\sigma= 84.65 $ km/s  (Ferrarese \& Merritt \cite{fm00})
for the three density profiles. Here, only DM contribution to the 
total mass is taken into account. 

We then assume that the total mass of the degenerate core should be between
$3\times 10^{3}M_{\odot}$ and $3.5 \times 10^{6} M_{\odot}$.
There are two arguments for the choice of this lower limit on the mass of fermion ball. 
First, this mass is needed
to boost the growth of stellar seed  black holes to
$ \sim 10^{6}M_{\odot}$ which then grow to $\sim 10^{9}M_{\odot}$ supermassive black hole
in SDSS quasars by baryonic accretion (Munyaneza \& Biermann \cite{mb05}).
Secondly, this is the  threshold mass,
below which no black hole will  be formed as in 
the case of M33 (Gebhardt et al. \cite{geb01}, \& Barth et al. \cite{barth05}). 

\begin{figure}
\includegraphics[width=8.5cm,height=4.5cm]{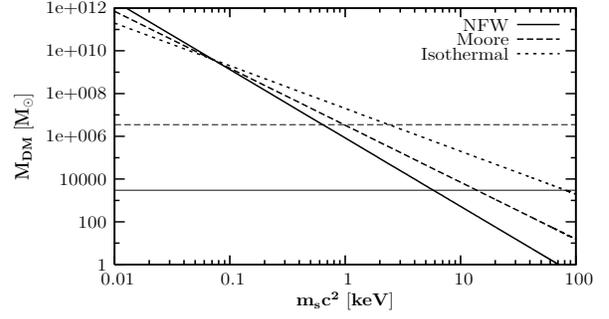}
\caption{The total mass $M_{DM}$ of the  fermion ball as a function of the
 fermion mass $m_{s}$. 
The total mass $M_{DM}$ scales with the the DM particle mass $m_{s}$ as $m_{s}^{-16/5}$,
$m_{s}^{-8/3}$ and as $m_{s}^{-2}$
for NFW, Moore and isothermal power law, respectively.
 Two horizontal lines 
at $M_{DM}=3.5 \times 10^{6} M_{\odot}$ and $M_{DM}=3 \times 10^{3}M_{\odot}$ have been drawn 
 to get the lower and upper limits on the mass of the DM particle. 
 }
\label{Fig.1}
\end{figure}
For NFW profile, the mass of the DM particle should be 
in the range 
of $ \ 0.6 \  {\rm keV} \stackrel {\textstyle <}{\sim} m_{s} \stackrel {\textstyle <}{\sim} 6 \ {\rm keV}$. A range
of  $ \ 1 \ {\rm keV} \stackrel {\textstyle <}{\sim} m_{s} \stackrel {\textstyle <}{\sim} 14 \ {\rm keV}$ is found
for Moore profile and finally we obtain 
$2 \ {\rm keV} \stackrel {\textstyle <}{\sim} m_{s} \stackrel {\textstyle <}{\sim} 82 \ {\rm keV}$
for the isothermal gas sphere. These limits on the  DM particle mass are illustrated in Fig.1 where
we plot the total mass of the degenerate core as a function of the DM mass $m_{s}$.
Using these lower limits for the DM particle, we predict
a density of $(1-100) \ M_{\odot}{\rm pc^{-3}}$ in  the case of NFW and Moore profiles
at the size of  a few parsec, which corresponds to
the values obtained in Klypin et al. (\cite{klypin02}).
In the case of the isothermal gas sphere, we predict a much higher density of a few 
$10^{5} M_{\odot} {\rm pc}^{-3}$
at a size of  about 1 {\rm pc}. 

Here, we want to point out that the total mass of DM core in the isothermal
gas sphere case is increased by a factor of nine
if we  use the full numerical solutions of 
Poisson's equation for a self gravitating system of fermions (Munyaneza \& Biermann \cite{mb05}).
Thus, in order to get  a degenerate core of
 $3.5 \times 10^{6}M_{\odot}$ with a velocity dispersion of  $\sigma=84.65~{\rm km/s}$, 
we get a fermion mass of $m_{s} \approx 7~{\rm keV}$ instead of
lower limit of $m_{s} \sim 2 ~ {\rm keV}$.
 Using a fully developed distribution function such as the one given in 
equation (\ref{eq:fdm}) would result in further
changes of the obtained limits. However, we do not expect any major 
modifications since the number of particles 
in the tail is much smaller compared to the main 
body of the phase space distribution.

\section{Conclusion and Discussion}
We have determined the constraints on the DM mass under the 
assumption that the NFW, Moore or isothermal
density profiles  become degenerate near the center. Assuming that the mass of the 
degenerate fermion ball is between about
 $3 \times 10^{3}M_{\odot}$  and 
  $3.5\times 10^{6}M_{\odot}$,
we have found that the mass of the DM particle
should have a lower limit of about 7 keV in the isothermal power law case.
 The  above limit 
on the DM mass is obtained from a simple model
which does not take into account baryonic matter
and a proper model for DM with anisotropic distribution in phase space
(Evans \& An \cite{evans06};  Hansen \& Stadel \cite{hs06}) and baryonic matter would modify these limits. 
The obtained mass range is in full agreement with the Tremaine-Gunn lower 
bound of about 0.5 {\rm keV }on the mass of any fermionic
DM when applied to dwarf spheroidal galaxies (Tremaine \& Gunn \cite{tg79}).
The constraints obtained  on DM particle mass  overlap with the upper limit on sterile neutrino masses
of about $8 \ {\rm keV}$
to obey
the X-ray emission constraints from the Virgo cluster observations (Abazajian \cite{abaza05}).
In addition,  a sterile neutrino mass in the mass range
 of 2-20 keVs  was derived 
 to explain   
the high velocities up
to 1000 km/s experienced by pulsars at birth  in 
supernovae  (Kusenko \cite{kusenko04}; Barkovitch
et al. \cite{barko04} and Fuller et al.  \cite{fuller03}).
Moreover, the obtained constraints agree well with the  lower 
bound of $m_{s} \approx 2 \  {\rm keV}$ on the mass of sterile neutrinos
 obtained from the analysis
 of the Lyman- $\alpha$ forest data (Hansen et al. \cite{hansen02}; Viel et al. \cite{viel05},
 and Abazajian \cite{abaza05}).
Recently, Biermann \& Kusenko (\cite{bk06}) have shown that the decay of 
such a sterile neutrino
could help initiate star formation in the early Universe and the
 detection of such an X-ray line would confirm
whether such a sterile neutrino exists or not and this 
 could be done 
by observing the X-ray decay line
using XMM and Chandra satellites (Boyarski et al. \cite{boya06},
 Riemer-S$\o$rensen et al. \cite{riemer06}). 

To summarize, we find that NFW, Moore and isothermal  density profiles become 
 degenerate for fermion DM  particle of ${\rm keV}$ masses
at a size of the order  parsec. Assuming that supermassive black holes grow from
 degenerate fermion
cores of masses of a few $10^{3} M_{\odot}$ to $\sim 10^{6}M_{\odot}$ at galactic centers,
 we have found that the  sterile neutrino mass
should be in the range from 0.6~{\rm keV} to about 82~{\rm keV}, with an improved lower limit
of about 7 keV in the isothermal power law density profile.
As this range of sterile neutrino masses
 overlaps with other results on sterile neutrino masses discussed in this Letter,
  this leaves the 
 sterile neutrino
to be an excellent candidate for DM.
The formation and growth details of degenerate fermion balls is beyond the scope of this Letter and 
will be discussed in another paper.
\begin{acknowledgements}
We are grateful to Alex Kusenko and Raoul Viollier 
 for useful discussions at SNAC06 in Crans-Montana. 
 We also thank Mikhail Shaposhnikov for 
 discussions at the Marcell Grossman meeting (MG11) in Berlin.
 FM thanks the organisers of SNAC06 for the hospitality in Crans-Montana and 
he acknowledges the financial support from the 
 Alexander von Humboldt foundation.
The work of PLB is supported by the Pierre Auger grant
05 CU 5PD1/2 via DESY/BMF.
\end{acknowledgements}

\end{document}